# Occupancy Estimation Using Low-Cost Wi-Fi Sniffers

Paolo Galluzzi, Edoardo Longo, Alessandro E. C. Redondi, Matteo Cesana
DEIB, Politecnico di Milano
Milan, Italy
Email: {name.surname}@polimi.it

*Abstract*—Real-time measurements on the occupancy status of indoor and outdoor spaces can be exploited in many scenarios (HVAC and lighting system control, building energy optimization, allocation and reservation of spaces, etc.). Traditional systems for occupancy estimation rely on environmental sensors (CO2, temperature, humidity) or video cameras. In this paper, we depart from such traditional approaches and propose a novel occupancy estimation system which is based on the capture of Wi-Fi management packets from users' devices. The system, implemented on a low-cost ESP8266 microcontroller, leverages a supervised learning model to adapt to different spaces and transmits occupancy information through the MQTT protocol to a web-based dashboard. Experimental results demonstrate the validity of the proposed solution in four different indoor university spaces.

## I. INTRODUCTION

In the last few years, the problem of accurately determining how many people are present in a specific room or building has received a lot of attention from the research community. Occupancy information is used primarily for controlling the indoor lighting or Heat, Ventilation and Air Conditioning (HVAC) systems of a building. System control may be performed in real-time or through the creation of prediction models for daily optimizations, with consequent reported energy savings ranging from 30% up to 80%, according to several studies [1], [2], [3]. Beside energy efficiency, occupancy information can be exploited to monitor the quality of indoor living and to maintain a comfortable environment for its occupants. In some scenarios, occupancy information can be also exploited to provide services to the building administrators and occupants. As an example, keeping track of the real time occupancy of spaces is a fundamental building block for automatic monitoring systems to prevent overcrowding or to trigger alerts in case of anomalies. Also, in large company settings, knowing in real-time which offices or rooms are free may be very valuable to quickly find places where to have a meeting. Similarly, knowing how many people are present in the room devoted to lunch may be useful to plan the lunch break and avoid the infamous queues at the microwave. In universities, knowing which classrooms are not occupied can be very useful for students in search of a quiet place to study or work on a project alone or in group.

Traditional systems for occupancy estimation rely on environmental sensors (CO2, temperature, humidity) or video cameras. However, both approaches have associated pros and cons: on the one hand environmental sensors are a cheap solution although generally not very accurate. On the other hand, camera-based systems are much more precise, but they are generally costly to deploy and maintain as well as being problematic for what concerns privacy issues.

In this paper, we take a different approach by using a novel occupancy estimation technique based on low-cost Wi-Fi sniffers. The rationale of using such a sensing technique is based on the observation that nowadays most people own a Wi-Fi enabled device and carry it with themselves everywhere, every time. Even if not connected to a Wi-Fi networks, such devices constantly transmit probe requests management frames, used to gather information about the available networks in range. Such frames are transmitted in the clear, and can therefore be easily captured by low-cost sniffers. Moreover, since each Probe Request frame contains the source MAC address of the transmitting device, processing such frames allows to estimate the number of distinct devices which are present around the sniffer, ultimately providing a measure of the occupancy status of a space.

Starting from this idea, in this paper we present a system able to perform occupancy estimation starting from data collected by low-cost sniffers. The system captures Probe Request packets and uses a supervised machine learning model to infer how many people are present in a certain area. The learning model is continuously adapted to the environment through supervised information, which are then feedback to the sniffers through the MQTT protocol. Finally, the occupancy estimation is delivered to a central server which provides real-time monitoring of the spaces of a building.

The paper is structured as it follows: Section II presents an overview of the state of the art concerning occupancy detection, with reference to different methods of estimation. Section III presents the implementation of the system, including the hardware design of the sniffer, the learning and estimation model as well as the data visualization service. Section IV evaluate the proposed system with experiments conducted in indoor spaces of a university department, and finally Section V concludes the paper.

## II. RELATED WORK

Several works in the last few years have tackled the problem of estimating the occupancy in indoor and outdoor spaces. The existing works can be categorized based on the type of sensing technology used in three main classes: (i) environmental-based, (ii) video-based and (iii) radio-based.

## A. Environmental sensing-based occupancy estimation

The majority of the works leverage environmental data to perform occupancy estimation. For what concerns indoor spaces, the gold standard input measurement used in this kind of works is the CO2 level, which is a good indicator of the number of person in a room and at the same time is able to preserve the privacy of the occupants [4]. Occupancy information is generally retrieved by analyzing the gradient of the CO2 level [5], or by solving the air mass balance equation [6]. However, both methods assume that the indoor CO2 concentration is uniform. Therefore, in real applications, these methods suffer from common issues such as unpredictable opening of doors and windows and uncertainties involved with the CO2 concentration level or its gradient, which lead to poor estimation accuracy. To improve the performance of such sensing technique, machine learning-based solutions can be used [7], additionally taking into account other environmental sources of information such as light, temperature and humidity [8], [9].

## B. Video-based occupancy estimation

A different approach consists in using video cameras for estimating the number of people occupying a space through image processing techniques [10]. For privacy considerations, cameras are generally installed in public spaces (hall, entrance) and attached at the ceiling, or very low image resolutions are used [11]. Other approaches ensuring privacy use cameras in the non-visible domain, such as passive infrared (PIR) sensors [12], [13] or depth cameras [14]. Compared to environmental sensing-based solutions, video-based approach allows for higher estimation accuracy, but are more costly to setup and maintain as well as being sensible to lighting conditions. For this reasons, some works propose hybrid data-fusion systems in which cameras are coupled with CO2 and other environmental sensors [15].

## C. Radio-based occupancy estimation

Very recently, some attention has been given to occupancy estimation systems which are based on radio measurements rather than on traditional sensors. Beside being very cheap, such systems have the benefit of being able to work both in indoor and outdoor spaces. The work in [16] uses a couple of transmitter and receiver devices to assess the impact of a certain number of people on the signal strength indicator at the receiver (RSS) on blocking the line of sight (LOS). Authors develop a model for the probability distribution of the received signal amplitude as a function of the total number of occupants and use that as estimation methods. Experiments in indoor spaces with up to 9 people allows for an average error below 2 people 95% of the times. The work in [17] uses as input data the Doppler spectrum of the Channel State Information (CSI) gathered at a single Wi-Fi receiver, rather than the RSS, with overall similar accuracy. Finally, in [18] the two approaches are compared on a single set of experiments. It is important to mention that the approach mentioned so far perform occupancy estimation with radio measurements without any assumption on the devices carried by people. However, the pervasiveness of Wi-Fi enabled devices allow to set up sensing systems which are tailored to a class of signals emitted by such devices, known as probe request frames. Such frames are broadcasted in-the-clear by personal devices for gathering information about the available networks and can be captured easily with a Wi-Fi sniffer [19]. Research studies on probe requests sniffing and analysis targeting user tracking, device classification, social analysis and privacy issues can be found in [20], [21], [22], [23]. To the best of our knowledge this is the first paper tackling the problem of precise occupancy detection using probe requests sniffing.

## III. PROPOSED SYSTEM

This section presents the proposed system. First, we describe the sensing process and the model used for estimating occupancy information. Then, we describe the hardware setup used and the software service built in order to support real-time monitoring of indoor space.

### A. Probe Requests sniffing

Capturing probe requests frames is as simple as receiving any other Wi-Fi frame and can be therefore achieved with any IEEE 802.11 compliant receiver set in *monitor* mode and listening on a single specific Wi-Fi channel[1]. Each received probe request allows to obtain several information, two of which are the most relevant for this work: the source device MAC address and the received signal strength (RSS) indicator. The source MAC address is a 48-bit string which identifies the device transmitting the probe request and whose first 3 bytes contain the Organizationally Unique Identifier (OUI) which identifies the radio chip vendor. The RSS is estimated at the receiver and is primarily related to the device output power and to the distance between the device and the receiver. These 2 pieces of data are enough for performing occupancy estimation: a sniffer device collects probe request frames for a certain amount of time $t$ (e.g., 5-10 minutes), counts the number of unique MAC addresses observed whose RSS is higher than a certain threshold and returns its estimate. Although very easy to implement, such system must cope with several issues:

1) *MAC randomization:* to avoid device tracking, some manufacturers transmit probe requests with a randomized, bogus MAC address. It is therefore important to distinguish valid MAC addresses from randomized ones to estimate occupancy correctly and avoid overestimation. In this paper, we use the OUI part of the MAC address to check if the address is authentic or not by looking up the OUI in the IEEE table of registered

---

[1]Wi-Fi devices transmit bursts of probe requests on all channels to gather information on all available networks. Listening on a single channel is therefore enough for capturing such frames

vendors. If the OUI is not in the table, the MAC address is considered as randomized.

2) *RSSI threshold:* ideally, one would like to capture probe requests only from the devices occupying the space under consideration. However, a sniffer is able to capture probe requests from all devices in its communication range, which for Wi-Fi may range from about 50 to 150 meters. It is hence important to reduce the communication range of the sniffer to match the boundary of the space under analysis. This can be conveniently done by establishing a RSS threshold and considering only probe requests with a RSS higher than such a threshold, which should be adapted according to the area under consideration.

3) *Device-user mapping:* finally, although is very common nowadays to always carry one Wi-Fi enabled device, this may not be always true. Some people may not carry a device, or have a device with its wireless interface off, or even have more than one Wi-Fi device. The number of unique MAC addresses observed from the captured probe requests cannot be therefore mapped directly to the number of occupants in a space.

To cope with these issues, we propose to use a supervised learning model that is able to adapt to the particular area under consideration and to its specific conditions.

### B. Occupancy estimation model

Let $N_V^\theta(t)$ the number of unique valid MAC addresses observed by a sniffer during a measurement time $t$ and whose RSS is higher than a threshold $\theta$. Similarly, let $N_R^\theta(t)$ be the number of unique random MAC addresses observed during the same measurement time. We propose to perform occupancy estimation using the following linear model:

$$\hat{N}(t) = \alpha N_V^\theta(t) + \beta N_R^\theta(t) \quad (1)$$

where $\alpha$ and $\beta$ are correction factors used to deal with the device-user mapping issue, and $\theta$ is the RSSI threshold. In order to learn $\alpha$, $\beta$ and $\theta$, we use a supervised learning approach. Ground truth occupancy values $N(t)$, $t = 1 \ldots T$ are provided to the system when available. The system then updates the parameters according to the following:

$$\alpha, \beta, \theta = \arg \min \frac{1}{T} \sum_{t=1}^{T} (\alpha N_V^\theta(t) + \beta N_R^\theta(t) - N(t))^2 \quad (2)$$

### C. Back-end system

The back-end system consists in a low-cost hardware sensing system and a backend software implemented as a web service. The sensing system is implemented using a NodeMCU ESP8266 microcontroller, a 3$ device which features a 32-bit RISC microprocessor running at 80 MHz, 16 KiB of RAM, several I/O interfaces (GPIO, SPI, I2C) and an IEEE 802.11 b/g/n compliant radio chipset supporting monitor mode. A BME280 sensor providing humidity, pressure and temperature information and a light sensor are also attached to the ESP8266

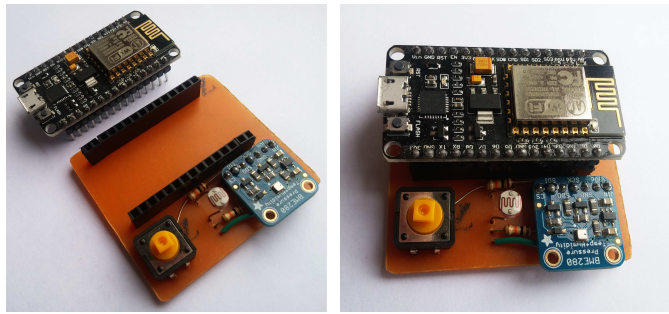

Fig. 1. The proposed ESP8266-based occupancy sensor. The attached BME280 sensor is used to enrich the occupancy information with environmental data from the monitored space. The yellow button is used to force the ESP8266 to stop sniffing probe requests and retrieve ground truth information from the server

(see Figure 1). At start up, the ESP8266 starts capturing probe requests: for each captured frame, it looks up in a locally cached copy of the IEEE table of registered OUI[2] if the probe request OUI is present or not. Table lookup is performed efficiently with indexing and takes only a couple of milliseconds. The ESP8266 also keeps a register of counters $N_V^{\theta_1}, N_V^{\theta_2}, \ldots, N_V^{\theta_M}$, where $\theta_1, \ldots, \theta_M$ are obtained by sampling the space of RSS values from -120 dBm to -44 dBm with $M = 40$. Upon reception of a probe request frame, its RSS $r$ is evaluated and the $N_V$ or $N_R$ counters are updated depending on the outcome of the table lookup. In details, all counters such that $\theta_i < r$ are incremented: as an example, $N_V^{\theta_1}$ counts all probe requests whose RSS is between -120 dBm and -118 dBm, and so on. Periodically (or when the yellow button in Figure 1 is pressed) the ESP8266 stops capturing probe requests and connects to the backend server using Wi-Fi for gathering ground truth information. If an updated ground truth measurement $N(t)$ is available, the ESP8266 fetches it and updates the parameters $\alpha, \beta, \theta$ by solving (2). The solution is found using brute force search: to limit the number of combinations of parameters to be evaluated, $\alpha$ and $\beta$ are limited between 0.1 and 2 and discretized with a step of 0.1. Such limits are determined considering a reasonable number of devices that each person can carry. The total number of combinations to evaluate is therefore $20 \times 20 \times 40 = 16000$, which takes 3.5 seconds on the ESP8266. After having found the optimal parameters, the system stores them and use them to periodically estimate the occupancy information using (1). Due to memory limits, the ESP8266 is able to store only 40 registers arrays $N_V^{\theta_1}, N_V^{\theta_2}, \ldots, N_V^{\theta_M}$, i.e. $T$ in (2) is equal to 40. After having reached this limit, information update on the ESP8266 is performed according to a FIFO queue.

The backend server is implemented using Node-RED, a visual tool built on top of the Node.js server-side framework that has recently become very popular in the development of IoT applications, thanks to its flexibility in creating quick software prototypes. The server has two main functions: (i) receiving and storing the estimates from the ESP8266 and (ii) transmitting the ground truth information on the occupancy status (if available) to the microcontroller. The two functions are implemented relying on the MQTT publish/subscribe protocol: the backend server acts as a MQTT broker, publish ground truth information and receives occupancy information from the ESP8266 sensor. The ground truth information is accompanied with a time to live (TTL) field which is set by the administrator. Upon connection to the broker, the ESP8266 ignores the ground truth information if the TTL is expired.

[2]http://standards-oui.ieee.org/oui.txt

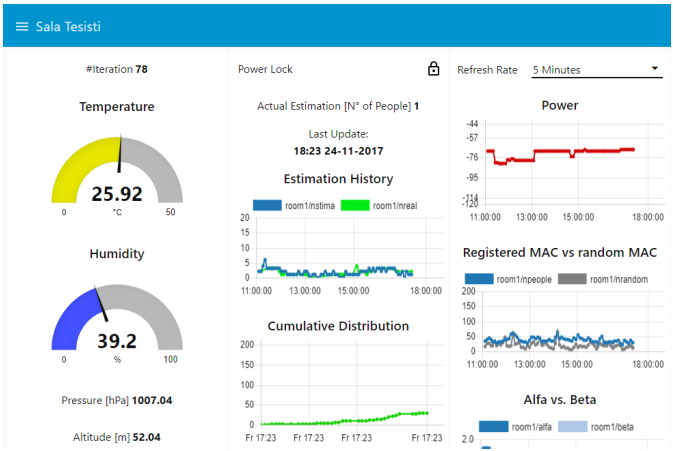

Fig. 2. Front end dashboard for occupancy monitoring. Environmental meauserements are also shown.

*D. Front-end system*

Leveraging the dashboard tools provided by Node-RED, a user interface is created with the purpose of easily accessing occupancy detection information and managing the system. A user can either access to a web-based graphical interface or use a Telegram chatbot to retrieve information on the occupancy status of a particular space. Beside the occupancy information estimated by the ESP8266, the dashboard (illustrated in Figure 2) also shows the environmental parameters retrieved by the sensors attached to the board, allowing a full monitoring of the space of interest.

## IV. Experiments

In order to evaluate the performance of the proposed system, we carried out measurements in 4 different rooms of an university building. The four rooms are characterised by different areas, number of seats and intended use. For each room, an ESP8266 occupancy sensor is deployed: periodically, ground truth information are provided to the sensor through the backend server by manually counting the number of people present in the room. Each time a ground truth information is available, the complete counter registers on board the ESP8266 are transmitted to the backend server. For each room, the process is repeated 200 times. The data collected on the backend serves is then divided into a training set and a test set. The training set consists of 40 ground truth / counters couples, mimicking the memory limits on the ESP8266, and it is used to estimate the optimal $\alpha$, $\beta$ and $\theta$ parameters. The remaining $P = 160$ measurements are used as a test set to evaluate the root mean squared error (RMSE) and the Mean Absolute Error (MAE) obtained by the proposed system, that is:

$$\text{RMSE} = \sqrt{\frac{1}{P} \sum_{i=1}^{P} (\alpha N_V^\theta(i) + \beta N_R^\theta(i) - N(i))^2} \quad (3)$$

$$\text{MAE} = \frac{1}{P} \sum_{i=1}^{P} |\alpha N_V^\theta(i) + \beta N_R^\theta(i) - N(i)| \quad (4)$$

The entire process is repeated 10 times according to $k$-fold crossvalidation, changing each time the set of 40 measurements used as training set and the corresponding test set. The final RMSE and MAE are averaged over the 10 validation folds.

Figure 3 shows the update process of the parameters ($\alpha$, $\beta$, $\theta$) as new training samples are provided to the system. As one can see, after an initial transient phase, the parameters of the model tend to converge around some stable values. Table I reports the result obtained for the four different rooms tested in the experiment. For each room we report the dimension and the number of seats available, the RMSE and MAE errors, the resulting $\alpha$, $\beta$ and RSS threshold as well as the percentage error computed with respect to the total capacity of the room. As one can see, the RMSE is limited between 2 and 4 people and the MAE between 2 and 8 people, depending on the size of the room. In general, the percentage error with respect to the number of seats available in a room is limited to about 10%. It can also be seen that the values of RSS threshold used are different in the four cases, and they do not seem to be correlated with the dimension of the room. The simple explanation for this values is that in crowded environments (such as those of the last two rooms analysed) the model prefer to keep the power threshold higher, but increase the $\alpha$ and $\beta$ values. On the contrary, in the first two rooms (which are in a more isolated scenario) the $\alpha$ and $\beta$ parameters are kept lower, while the power threshold (which determines the radius of action of the sensor) is lower. In general, the results obtained are comparable to what achieved by systems based on CO2 sensing [4], with the difference that the proposed system does not require strong assumption on the boundary conditions of the environment (number of windows/door open, air and ventilation flow parameters, etc.) and can be in principle used also for outdoor spaces.

## V. Conclusion

We proposed a system for estimating the occupancy of a space based on the capture of Wi-Fi probe requests messages and validated it through experimental results. The system is able to adapt to the particular space under analysis through a supervised learning model. A front-end dashboard is also provided to easily monitor the space. Future works (i) will target the development of advanced machine learning models to improve the occupancy detection estimation considering additional inputs coming from environmental sensors and (ii) will evaluate the performance also in an outdoor scenario.

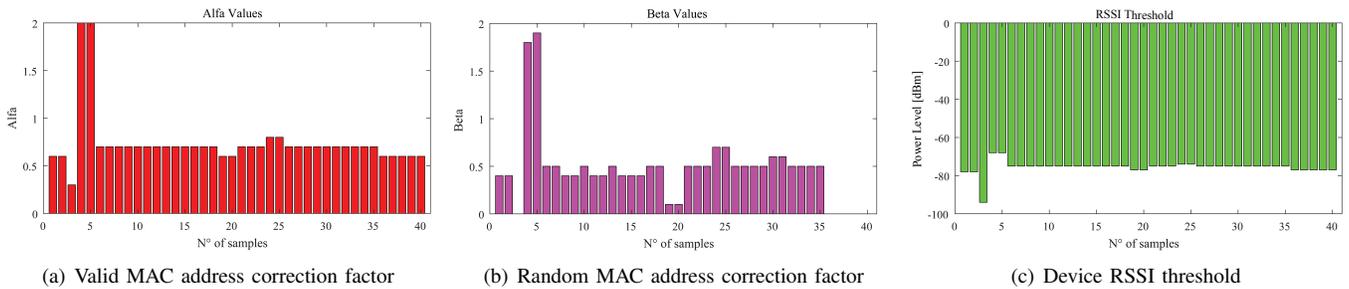

(a) Valid MAC address correction factor  (b) Random MAC address correction factor  (c) Device RSSI threshold

Fig. 3. Variables calculated in the ANTLAB room during the training part. The algorithm finds the value $\alpha$, $\beta$, $\theta$ that minimize the percentage error.

TABLE I
EXPERIMENTAL RESULTS

| Room | Area [m$^2$] | No of seats | $\alpha$ | $\beta$ | Rssi Threshold $\theta$ [dBm] | RMSE [ppl] | MAE [ppl] | Percentage Error [%] |
|---|---|---|---|---|---|---|---|---|
| **ANTLAB** | 100 | 21 | 0.54 | 0.05 | -79.45 | 2.14 | 2.23 | 10.6 |
| **PhD Students Room** | 21 | 18 | 0.71 | 0.15 | -79.3 | 1.61 | 1.72 | 10.75 |
| **Classroom B.5.3** | 60 | 80 | 1.5 | 1.03 | -55.8 | 4.14 | 7.8 | 9.75 |
| **Classroom L.26.01** | 54 | 70 | 1.3 | 0.1 | -59.23 | 2.04 | 4.37 | 6.24 |